\documentstyle[12pt,graphicx,epsf,braket]{article}
\begin{document}

\def\AEF{A.E. Faraggi}

\def\vol#1#2#3{{\bf {#1}} ({#2}) {#3}}
\def\NPB#1#2#3{{\it Nucl.\ Phys.}\/ {\bf B#1} (#2) #3}
\def\PLB#1#2#3{{\it Phys.\ Lett.}\/ {\bf B#1} (#2) #3}
\def\PRD#1#2#3{{\it Phys.\ Rev.}\/ {\bf D#1} (#2) #3}
\def\PRL#1#2#3{{\it Phys.\ Rev.\ Lett.}\/ {\bf #1} (#2) #3}
\def\PRT#1#2#3{{\it Phys.\ Rep.}\/ {\bf#1} (#2) #3}
\def\MODA#1#2#3{{\it Mod.\ Phys.\ Lett.}\/ {\bf A#1} (#2) #3}
\def\RMP#1#2#3{{\it Rev.\ Mod.\ Phys.}\/ {\bf #1} (#2) #3}
\def\IJMP#1#2#3{{\it Int.\ J.\ Mod.\ Phys.}\/ {\bf A#1} (#2) #3}
\def\nuvc#1#2#3{{\it Nuovo Cimento}\/ {\bf #1A} (#2) #3}
\def\RPP#1#2#3{{\it Rept.\ Prog.\ Phys.}\/ {\bf #1} (#2) #3}
\def\APJ#1#2#3{{\it Astrophys.\ J.}\/ {\bf #1} (#2) #3}
\def\APP#1#2#3{{\it Astropart.\ Phys.}\/ {\bf #1} (#2) #3}
\def\EJP#1#2#3{{\it Eur.\ Phys.\ Jour.}\/ {\bf C#1} (#2) #3}
\def\etal{{\it et al\/}}

\newcommand{\cc}[2]{c{#1\atopwithdelims[]#2}}
\newcommand{\bev}{\begin{verbatim}}
\newcommand{\beq}{\begin{equation}}
\newcommand{\beqa}{\begin{eqnarray}}
\newcommand{\beqn}{\begin{eqnarray}}
\newcommand{\eeqn}{\end{eqnarray}}
\newcommand{\eeqa}{\end{eqnarray}}
\newcommand{\eeq}{\end{equation}}
\newcommand{\beqt}{\begin{equation*}}
\newcommand{\eeqt}{\end{equation*}}
\newcommand{\Eev}{\end{verbatim}}
\newcommand{\bec}{\begin{center}}
\newcommand{\eec}{\end{center}}
\def\ie{{\it i.e.}}
\def\eg{{\it e.g.}}
\def\half{{\textstyle{1\over 2}}}
\def\nicefrac#1#2{\hbox{${#1\over #2}$}}
\def\third{{\textstyle {1\over3}}}
\def\quarter{{\textstyle {1\over4}}}
\def\m{{\tt -}}
\def\mass{M_{l^+ l^-}}
\def\p{{\tt +}}

\def\slash#1{#1\hskip-6pt/\hskip6pt}
\def\slk{\slash{k}}
\def\GeV{\,{\rm GeV}}
\def\TeV{\,{\rm TeV}}
\def\y{\,{\rm y}}

\def\l{\langle}
\def\r{\rangle}
\def\LRS{LRS  }

\begin{titlepage}
\samepage{
\setcounter{page}{1}
\rightline{LTH--917} 
\vspace{1.5cm}
\begin{center}
 {\Large \bf Leptophobic $Z^\prime$ in\\ Heterotic--String Derived Models}
\vspace{.25 cm}

Alon E. Faraggi\footnote{
		                  E-mail address: faraggi@amtp.liv.ac.uk}
and 
Viraf M. Mehta\footnote{ E-mail address:
	                          Viraf.Mehta@liv.ac.uk}
\\
\vspace{.25cm}
{\it Department of Mathematical Sciences\\
University of Liverpool, Liverpool, L69 7ZL, United Kingdom}
\end{center}

\begin{abstract}
The CDF collaboration's recent observation of an excess of events 
in the $Wjj$ channel may be attributed to a new Abelian vector boson
with suppressed couplings to leptons. While D0 finds no evidence
of an excess, the CDF data provide an opportunity to revisit an old result 
on leptophobic $Z^\prime$ in heterotic--string derived models. 
We re-examine the conditions for the existence of a leptophobic 
$U(1)$ symmetry, which arises from a combination of the 
$U(1)_{B-L}$ symmetry and the horizontal flavour symmetries,
to form a universal $U(1)$ symmetry. While the conditions for 
the existence of a leptophobic combination  are not generic, 
we show that the left--right symmetric free fermionic
heterotic--string models also admit a leptophobic combination. 
In some cases the leptophobic $U(1)$ is augmented by the
enhancement of the colour group, along the lines 
of models proposed by Foot and Hernandez. 

\end{abstract}
\smallskip}
\end{titlepage}

The discrepancy between the recent CDF \cite{cdf} and D0 \cite{d0} 
results suggests considerable ambiguity as to whether there
is an excess of $Wjj$ events in the $Mjj\sim140{\rm GeV}$ region. 
Nevertheless, it is interesting to explore various theoretical
scenarios that can account for such an excess. 
Indeed, various proposals appeared in the literature 
to explain the CDF results within and beyond the Standard Model
\cite{SMapproaches,zprimeapproaches, typeIstring} . 
One proposal amongst those attributes the
discrepancy to a new Abelian vector boson in the appropriate 
mass range \cite{zprimeapproaches,typeIstring}.
However, since an enhancement in the dilepton
channel is not observed, as well constraints 
arising from direct production at LEPII, Tevatron and LHC searches,
the leptonic couplings of a putative $Z^\prime$ have to be suppressed. 

Additional Abelian space--time vector bosons beyond those that mediate
the $SU(3)\times SU(2)\times U(1)_Y$ subatomic interactions are abundant
in extensions of the Standard Model \cite{zprimeliterature}. Indeed, they
arise in Grand Unified theories, based on $SO(10)$ and $E_6$ 
gauge extensions of the Standard Model gauge group,
which are well motivated by the Standard Model matter 
states and charges. Similarly, Abelian extensions of the 
Standard Model are common in string theories. However,
most of these extensions will produce extra bosons 
with unsuppressed coupling to leptons. It is therefore 
of interest to examine how a leptophobic $Z^\prime$ can arise 
\cite{leptophobic, typeIstring}. 
Obviously, one can simply gauge the baryon number $U(1)_B$, and
this exercise has been undertaken, \cite{perezwise},
and within type I string theories a gauged $U(1)_B$ may indeed arise. 
However, in Grand Unified theories, as well as in an heterotic--string theory
that accommodates them, Abelian extensions of the Standard Model typically
have unsuppressed couplings to leptons. 

One exception to this generic expectation was the 
heterotic--string model of ref. \cite{leptophobic}.
The recent CDF data provide an opportune moment to re-examine
how a leptophobic $Z^\prime$ can arise in heterotic--string models.
In this respect the type I and heterotic--string cases imply different
phenomenological signatures beyond the leptophobic $Z^\prime$
that will be instrumental in distinguishing between the 
two cases. While the heterotic--string maintains
the Grand Unified embedding 
of the Standard Model states, the type I string 
does not. In particular, the heterotic--string can still 
preserve the embedding of the Standard Model matter states
in spinorial 16 representations of $SO(10)$, 
which is well motivated by the Standard Model data. 
In the type I scenario the string scale is lowered to the
TeV scale, which will be signalled by the emergence of Regge recurrences
at parton collision energies $\sqrt{\hat s}\sim M_s\equiv {\rm string~scale}$.
In the heterotic case the string scale is still at the 
Planck scale. The big desert between the weak and Planck scales is
preserved, albeit with an unexpected oasis in between. 

In this paper we therefore re-examine the ingredients that produced
the leptophobic $Z^\prime$ model of ref. \cite{leptophobic}. 
The main feature of this model is that the $U(1)_{B-L}$ gauge symmetry,
which is embedded in $SO(10)$, plus a combination of the flavour 
$U(1)$ symmetry produces a family universal, leptophobic $U(1)$ symmetry.
The additional $U(1)$ symmetries compensate for the lepton number
in $U(1)_{B-L}$ and the resulting $U(1)$ therefore becomes a
gauged baryon number. In the specific model of ref. \cite{leptophobic}
the colour gauge symmetry is enhanced from $SU(3)_C\times U(1)_B$ to 
$SU(4)_C$, due to space--time vector bosons that arise from twisted
sectors. We discuss how leptophobic $U(1)$ symmetries may arise 
in this class of superstring compactifications without enhancement
of the gauge group. In particular, we show that the class 
of left--right symmetric models of ref. \cite{lrs} 
reproduces the conditions that admits a leptophobic 
$U(1)$ combination without gauge enhancement. 

The superstring models that we discuss are constructed in the 
free fermionic formulation \cite{fff}. 
In this formulation a model is constructed by 
choosing a consistent set of boundary condition basis vectors.
The basis vectors, $b_k$, span a finite  
additive group $\Xi=\sum_k{{n_k}{b_k}}$
where $n_k=0,\cdots,{{N_{z_k}}-1}$.
The physical massless states in the Hilbert space of a given sector
$\alpha\in{\Xi}$, are obtained by acting on the vacuum with 
bosonic and fermionic operators and by
applying the generalised GSO projections. The $U(1)$
charges, $Q(f)$, with respect to the unbroken Cartan generators of the four 
dimensional gauge group, which are in one 
to one correspondence with the $U(1)$
currents ${f^*}f$ for each complex fermion f, are given by:
\begin{equation}
{Q(f) = {1\over 2}\alpha(f) + F(f)},
\label{u1charges}
\end{equation}
where $\alpha(f)$ is the boundary condition of the world--sheet fermion $f$
 in the sector $\alpha$, and 
$F_\alpha(f)$ is a fermion number operator counting each mode of 
$f$ once (and if $f$ is complex, $f^*$ minus once). 
For periodic fermions,
$\alpha(f)=1$, the vacuum is a spinor in order to represent the Clifford
algebra of the corresponding zero modes. 
For each periodic complex fermion $f$
there are two degenerate vacua, ${\vert +\rangle},{\vert -\rangle}$ , 
annihilated by the zero modes $f_0$ and
${{f_0}^*}$ and with fermion numbers  $F(f)=0,-1$ respectively. 

The realistic models in the free fermionic formulation are generated by 
a basis of boundary condition vectors for all world--sheet fermions 
\cite{fsu5,fny,alr,eu,mshsm,lrs,exophobic}. 
The basis is constructed in two stages. The first stage consist
of the NAHE set \cite{nahe}, 
which is a set of five boundary condition basis 
vectors, $\{{{\bf 1},S,b_1,b_2,b_3}\}$. The gauge group after the NAHE set
is $SO(10)\times SO(6)^3\times E_8$ with $N=1$ space--time supersymmetry. 
The space--time vector bosons that generate the gauge group 
arise from the Neveu--Schwarz (NS) sector and from the sector $1+b_1+b_2+b_3$. 
The Neveu--Schwarz sector produces the generators of 
$SO(10)\times SO(6)^3\times SO(16)$. The sector $1+b_1+b_2+b_3$
produces the spinorial 128 of $SO(16)$ and completes the hidden 
gauge group to $E_8$. The vectors $b_1$, $b_2$ and $b_3$ 
produce 48 spinorial 16 of $SO(10)$, sixteen from each sector $b_1$, 
$b_2$ and $b_3$. The vacuum of these sectors contains eight right--moving  periodic 
fermions. Five of those periodic fermions produce the charges under the 
$SO(10)$ group, while the remaining three periodic fermions 
generate charges with respect to the flavour symmetries. Each of the 
sectors $b_1$, $b_2$ and $b_3$ is charged with respect to a different set 
of flavour quantum numbers, $SO(6)_{1,2,3}$.

The NAHE set divides the 44 right--moving and 20 left--moving real internal
fermions in the following way: ${\bar\psi}^{1,\cdots,5}$ are complex and
produce the observable $SO(10)$ symmetry; ${\bar\phi}^{1,\cdots,8}$ are
complex and produce the hidden $E_8$ gauge group;
$\{{\bar\eta}^1,{\bar y}^{3,\cdots,6}\}$, $\{{\bar\eta}^2,{\bar y}^{1,2}
,{\bar\omega}^{5,6}\}$, $\{{\bar\eta}^3,{\bar\omega}^{1,\cdots,4}\}$
give rise to the three horizontal $SO(6)$ symmetries. The left--moving
$\{y,\omega\}$ states are divided into, $\{{y}^{3,\cdots,6}\}$, $\{{y}^{1,2}
,{\omega}^{5,6}\}$, $\{{\omega}^{1,\cdots,4}\}$. The left--moving
$\chi^{12},\chi^{34},\chi^{56}$ states carry the supersymmetry charges.
Each sector $b_1$, $b_2$ and $b_3$ carries periodic boundary conditions
under $(\psi^\mu\vert{\bar\psi}^{1,\cdots,5})$ and one of the three groups:
$(\chi_{12},\{y^{3,\cdots,6}\vert{\bar y}^{3,\cdots6}\},{\bar\eta}^1)$,
$(\chi_{34},\{y^{1,2},\omega^{5,6}\vert{\bar y}^{1,2}{\bar\omega}^{5,6}\},
{\bar\eta}^2)$, 
$(\chi_{56},\{\omega^{1,\cdots,4}\vert{\bar\omega}^{1,
\cdots4}\},{\bar\eta}^3)$. 

The division of the internal fermions is a
reflection of the underlying $Z_2\times Z_2$ orbifold 
compactification \cite{xmap}. 
The Neveu--Schwarz sector corresponds to the
untwisted sector and the sectors $b_1$, $b_2$ and $b_3$ correspond to the
three twisted sectors of the $Z_2\times Z_2$ orbifold models.
At this level there is a discrete $S_3$ 
permutation symmetry between the three sectors $b_1$, $b_2$ 
and $b_3$. This permutation symmetry arises due to the symmetry of the 
NAHE set and may be essential for the universality of the leptophobic
$U(1)$ symmetry. Due to the underlying $Z_2\times Z_2$ orbifold
compactification, each of the chiral generations from the sectors 
$b_1$, $b_2$ and $b_3$ is charged with respect to a different set 
of flavour charges. 

The second stage of the basis construction consists of adding three
additional basis vectors to the NAHE set.
Three additional vectors are needed to reduce the number of generations 
to three; one from each sector $b_1$, $b_2$ and $b_3$. 
One specific example is given in table 1. 
The choice of boundary
conditions to the set of real internal fermions
${\{y,\omega\vert{\bar y},{\bar\omega}\}^{1,\cdots,6}}$  
determines the low energy properties, like the number of generations,
Higgs doublet--triplet splitting and Yukawa couplings \cite{pstudies}.
 
The final gauge group in the free fermionic standard--like models arises as follows. 
The Neveu--Schwarz sector produces the 
generators of $SU(3)_C\times SU(2)_L\times U(1)_C\times U(1)_L
\times U(1)_{1,2,3}\times U(1)_{4,5,6}\times {hidden}$, 
where the $hidden$ gauge group arises from the hidden $E_8$ gauge group of
the heterotic--string in ten dimensions. 
The $SO(10)$ symmetry is broken to 
$SU(3)_C\times U(1)_C\times SU(2)_L\times U(1)_L$\footnote{
      			 $U(1)_C={3\over2}U(1)_{B-L}$ and
      			 $U(1)_L=2U(1)_{T_{3_R}}.$}, where
\begin{eqnarray}
U(1)_C&&={\rm Tr} U(3)_C~\Rightarrow~Q_C=
			\sum_{i=1}^3Q({\bar\psi}^i), \nonumber\\
U(1)_L&&={\rm Tr} U(2)_L~\Rightarrow~Q_L=
			\sum_{i=4}^5Q({\bar\psi}^i).
\label{u1cl}
\end{eqnarray}
The flavour $SO(6)^3$ symmetries are broken to $U(1)^{3+n}$ with
$(n=0,\cdots,6)$. The first three, denoted by $U(1)_{j}$, arise 
from the world--sheet currents ${\bar\eta}^j{\bar\eta}^{j^*}$
$(j=1,2,3)$. These three $U(1)$ symmetries are present in all
the three generation free fermionic models which use the NAHE set. 
Additional horizontal $U(1)$ symmetries, denoted by $U(1)_{j}$ 
$(j=4,5,...)$, arise by pairing two real fermions from the sets
$\{{\bar y}^{3,\cdots,6}\}$, 
$\{{\bar y}^{1,2},{\bar\omega}^{5,6}\}$, and
$\{{\bar\omega}^{1,\cdots,4}\}$. 
The final observable gauge group depends on
the number of such pairings. In the model of ref. \cite{leptophobic} there are 
three such pairings, ${\bar y}^3{\bar y}^6$, ${\bar y}^1{\bar\omega}^5$
and ${\bar\omega}^2{\bar\omega}^4$, which generate three additional 
$U(1)$ symmetries, denoted by $U(1)_{{4,5,6}}$. It is 
important to note that the existence of these three additional 
$U(1)$ currents is correlated with a superstring doublet--triplet
splitting mechanism \cite{pstudies}. Due to these extra $U(1)$ symmetries, 
the colour triplets from the NS sector are projected out of the spectrum 
by the GSO projections while the electroweak doublets remain in the 
light spectrum. 

The key to understanding how the leptophobic $U(1)$ arises in the model of ref. \cite{leptophobic}
are the charges of the matter states from the sectors $b_1$, $b_2$ and $b_3$ under the 
flavour $U(1)_{j}$ with $j=4,5,6$. For example, the charges of the states from the sector $b_1$ 
are: 
\begin{eqnarray}
&&({e_L^c}+{u_L^c})_{{1\over2},0,0,{1\over2},0,0}+\nonumber\\
&&({d_L^c}+{N_L^c})_{{1\over2},0,0,{{1\over2}},0,0}+\nonumber\\
&&(L)_{{1\over2},0,0,-{1\over2},0,0}+(Q)_{{1\over2},0,0,-{1\over2},0,0},
\label{lpmodelb1}
\end{eqnarray}
and similarly for the states from the sectors $b_2$ and $b_3$. 
With these charge assignments, the quarks are charged with respect to 
the following combination
\begin{equation}
U(1)_B={1\over3}U_C-(U_{r_4}+U_{r_5}+U_{r_6}),
\label{u1b}
\end{equation}
whereas the leptons are neutral with respect to it. Hence, this combination is a family universal,
leptophobic $U(1)$ symmetry. In the model of ref. \cite{leptophobic} 
additional space--time vector bosons arise from the sector $X=1+\alpha+2\gamma$
in which $X_L\cdot X_L=0$ and $X_R\times X_R=8$. The additional vector bosons 
transform as triplets of $SU(3)_C$ and enhance it to $SU(4)_C$, where  the $U(1)$ combination given 
by $$U(1)_{B^\prime}=U(1)_B-U_7+U_9,$$
is the $U(1)$ generator of the enhanced $SU(4)$ symmetry. Here, $U_7$ and $U_9$ arise 
from the world--sheet complex fermions $\bar\phi^1$ and $\bar\phi^8$. The full massless spectrum
and charges of this model were given in ref. \cite{leptophobic}.  In this model the 
$U(1)_{1,2,3}$ symmetries are anomalous with ${\rm Tr}U_1=24$, ${\rm Tr}U_2=24$ and
${\rm Tr}U_3=24$. Hence, the family universal combination of these three $U(1)$ is
anomalous, whereas the two family non--universal combinations are anomaly free. 
The $U(1)_{4,5,6,7,9}$, as well as $U(1)_{B-L}$, are, however, anomaly free.
Hence, the leptophobic $U(1)$ combination is anomaly free and can remain, 
in principle, unbroken down to low scales.    

The existence of a leptophobic, family universal and anomaly free $U(1)$ is highly non--trivial and 
not generic in phenomenological heterotic--string models. To demonstrate that this is 
indeed the case, we examine the model of \cite{eu}. 
The sectors $b_{1,2,3}$ 
produce the three chiral generations that are charged with respect to the 
same flavour symmetries, but differ from the corresponding charges in the model
of ref. \cite{leptophobic}. For example, the states from the sector $b_1$ carry the
following charges:
\begin{eqnarray}
&&({e_L^c}+{u_L^c})_{{1\over2},0,0,{1\over2},0,0}+\nonumber\\
&&({d_L^c}+{N_L^c})_{{1\over2},0,0,{-{1\over2}},0,0}+\nonumber\\
&&(L)_{{1\over2},0,0,{1\over2},0,0}+(Q)_{{1\over2},0,0,-{1\over2},0,0}.
\label{eu1b1}
\end{eqnarray}
We observe that $e^c_L$ ad $L$ have like--sign charges under $U(1)_4$. Since they
carry opposite sign charges under $U(1)_C$, $U(1)_4$ cannot be used to cancel the $B-L$
charge for both 
these states. Since they carry like--sign charges under $U(1)_1$, a leptophobic, family universal 
$U(1)$ cannot be made from these $U(1)$ symmetries. 
The model of ref. \cite{eu} preserves the cyclic permutation of the NAHE set. 
Hence, a similar charge assignment is obtained in the sectors $b_2$ and $b_3$. 
In this model the flavour symmetries $U(1)_{4,,5,6}$ are anomalous. Therefore, 
their combination with $U(1)_C$ is not anomaly free and must be broken.

As a second negative example, we consider the model of ref. \cite{fny}. 
In this model the states from the sector $b_1$ carry the following $U(1)$ charges
\begin{eqnarray}
&&({e_L^c}+{u_L^c})_{-{1\over2},0,0,-{1\over2},0,0}+\nonumber\\
&&({d_L^c}+{N_L^c})_{-{1\over2},0,0,{-{1\over2}},0,0}+\nonumber\\
&&(L)_{-{1\over2},0,0,{1\over2},0,0}+(Q)_{-{1\over2},0,0,{1\over2},0,0}.
\label{fnyb1}
\end{eqnarray}
In this sector the combination given in eq. (\ref{u1b}) is leptophobic. 
However, the states from the sector $b_2$ have charges
\begin{eqnarray}
&&({e_L^c}+{u_L^c})_{0, -{1\over2},0,0,{1\over2},0}+\nonumber\\
&&({d_L^c}+{N_L^c})_{0, -{1\over2},0,0,{-{1\over2}},0}+\nonumber\\
&&(L)_{0,{1\over2},0,0,-{1\over2},0}+(Q)_{0,{1\over2},0,0,{1\over2},0},
\label{fnyb2}
\end{eqnarray}
Hence, in this sector the combination (\ref{u1b}) is not leptophobic and is not family
universal. Furthermore, the flavour symmetries are anomalous in this model and, 
consequently, as is the combination given in eq. (\ref{u1b}).

Is the existence of a leptophobic $U(1)$ combination therefore a peculiarity of
the model of ref. (\cite{leptophobic})? As seen from the charge assignments
in eq. (\ref{lpmodelb1}) the key is that the charges of the left-- and right--handed
fields differ in sign with respect to $U_{4,5,6}$ in the sectors $b_1$, $b_2$ and $b_3$,
respectively. This model preserves the cyclic permutation symmetry of the NAHE 
set and therefore, the $U(1)$ combination in eq. (\ref{u1b}), is family universal.
Furthermore, $U(1)_{4,5,6}$ are anomaly free in the model of ref. \cite{leptophobic} and
therefore, their combination with $U(1)_{B-L}$ is also anomaly free. In this model
the gauge symmetry is enhanced by space--time vector bosons arising from the twisted
sectors. However, we can envision a more systematic classification, along the lines
of ref. \cite{fkr,exophobic}, and that the extra bosons can be projected out from 
the spectrum in vacua that resemble the properties of this model. In such 
a case, the leptophobic $U(1)$ will arise without enhancement. 

As seen from the other two examples provided by the models in refs. \cite{eu} and 
\cite{fny}, the existence of a family universal, anomaly free leptophobic $U(1)$ 
combination in heterotic--string vacua is highly non--trivial. A class of models 
that reproduces the conditions for the existence of such a $U(1)$ combination
are the left--right symmetric models of ref. \cite{lrs}. However, in this case
the $U(1)$ symmetries that are combined with $U(1)_{B-L}$ are 
not the flavour $U(1)_{4,5,6}$, but rather the $U(1)_{1,2,3}$. This possibility is particular
to the left--right symmetric heterotic--string models \cite{lrs}, and is not applicable in
the other quasi--realistic free fermionic models, in which the $SO(10)$ symmetry
is broken to the flipped $SU(5)$, $SO(6)\times SO(4)$ or $SU(3)\times SU(2)\times U(1)^2$
subgroups. The reason is that, in these cases, the charges of all the states 
from a given sector $b_j$ are the same with respect to $U(1)_j$ with $j=1,2,3$. 
This situation arises because the states from the sectors $b_j$ in
these models preserve the $E_6$ charge assignment under the decomposition 
$E_6\rightarrow SO(10)\times U(1)$. A further consequence is that the $U(1)$ combination
which arises from $E_6$ becomes anomalous in  these models \cite{anomalousu1}.

On the other hand, in the left--right symmetric models, the GSO projection
that breaks the $SO(10)$ symmetry to $SU(3)\times U(1)\times SU(2)^2$ dictates
that the $U(1)_{1,2,3}$ charges of the left--handed fields, $Q_L$ and $L_L$,
differs in sign from those of the right--handed fields, $Q_R\equiv u_L^c+d_L^c$ and 
$L_R\equiv e_L^c+ N_L^c$. Their charges with respect to $U(1)_{4,5,6}$
may, or may not differ in sign. Hence, for example in the first model of ref. \cite{lrs},
we find for the sector $b_1$ 
\begin{eqnarray}
&&({u_L^c}+{d_L^c})_{ {1\over2},0,0,{1\over2},0,0}+\nonumber\\
&&({e_L^c}+{N_L^c})_{ {1\over2},0,0,{-{1\over2}},0,0}+\nonumber\\
&&(L)_{-{1\over2},0,0,{1\over2},0,0}+(Q)_{-{1\over2},0,0,-{1\over2},0,0},
\label{lrsm1b1}
\end{eqnarray}
with similar charges under $U(1)_{2,3}$ for the states from the sectors $b_2$ and $b_3$, 
respectively. The $U(1)$ combination given by
\beq
U(1)_B={1\over 3}U_C-U_1-U_2-U_3, 
\label{u1c123}
\eeq
is family universal, anomaly free and leptophobic. In the left--right 
symmetric models, the $U(1)_{1,2,3}$ are anomaly free due to the 
specific symmetry breaking pattern and consequent charge assignments, 
whereas $U(1)_{4,5,6}$ may be anomalous or anomaly free in different models. 
The left--right symmetric free fermionic heterotic--string models therefore provide 
a second example that produces a potentially viable leptophobic $U(1)$ at
low scales. In both cases, it is seen that the mechanism that yields a leptophobic 
$U(1)$ symmetry involves the existence of a combination of flavour $U(1)$ symmetries 
that nullifies the lepton number component of $U(1)_{B-L}$. The left--right 
symmetric models produce examples that are completely free of any gauge
or gravitational anomalies. Specifically, all $U(1)$ symmetries in these models are anomaly
free. Hence, any combination of the $U(1)$ symmetries, including the leptophobic 
combination, is anomaly free.

To guarantee that a $U(1)$ symmetry remains viable down to low scales,
we must ensure that the spectrum remains anomaly free down to these 
scales. If we just consider the Standard Model states $U(1)_B$ 
has various mixed anomalies, which are compensated by additional 
states that arise in the string models. This additional spectrum is 
highly model dependent, but is constrained by the string charge 
assignments. The issue of how the $U(1)$ symmetry can remain
viable down to low scales is, therefore, model dependent and 
highly non--trivial. In type I string theories, this is solved by
lowering the string scale down to the TeV scale. However, 
generically, one expects in this case, dangerous 
proton decay mediating operators to be generated (see, however, 
\cite{lebedmeyes} that suggests otherwise). This scenario,
in any case, has a distinct signature in the form of recurring 
Regge resonances, which will be confirmed or refuted in
forthcoming LHC experiments. From the point of view of a bottom--up 
approach, gauging baryon number is possible by
judicially adding states with appropriate charges. 
However, the top--down approach relies on the states 
and charges that are compatible with the string charge 
assignments and other constraints. Therefore, the states 
that are contemplated in the bottom--up approach are not likely to
exist in string constructions. Furthermore, string models typically produce
exotic fractionally charged states that are severely constrained by
experiments. String models in which the exotic states only appear in the 
massive spectrum do exist \cite{exophobic}. However, in these models 
the charge assignments are mundane. Recently, we have been able to 
construct effective field theories with an effective low scale $U(1)$ that 
suppresses proton decay mediating $U(1)$ \cite{viraf}. However, this $U(1)$ is not 
leptophobic. All in all, an interesting possibility is that the
enhanced non--Abelian symmetry in the model of ref. \cite{leptophobic}
is not superfluous, but required to maintain a viable leptophobic $U(1)$ 
down to the low scale. This scenario will then fall into the class
of models considered in ref. \cite{extendedcolor}, in which the colour
group is enhanced. It has also been proposed \cite{foot2011} that this class of 
theories may explain the top forward--backward asymmetry, 
which is indicated by the CDF experiment \cite{topbfasymmetry}.
While in the leptophobic model of ref. \cite{leptophobic} the colour group 
is enhanced to $SU(4)$, enhancement to $SU(5)$ is also possible if $SU(3)_C$
combines with an hidden $SU(2)$ group factor.  

In this paper we discussed how a leptophobic $U(1)$ symmetry may arise 
in heterotic--string derived models. The examples that we considered preserve
the embedding of the Standard Model matter states in spinorial 16 
representations of $SO(10)$. The leptophobic $U(1)$ arises from a combination of
the $U(1)_{B-L}$ symmetry, which is embedded in $SO(10)$, and the horizontal
flavour symmetries, which effectively cancels the lepton charge, 
resulting in a gauged baryon number. This may, or may not, be augmented 
by additional vector bosons that enhance the colour group.
If forthcoming data provides further weight to the CDF claims, rather than to D0,
the focus of model building will clearly shift in that direction. 

\section*{Acknowledgements}

AEF would to thank the theory division at CERN for hospitality. 
This work was supported in part by the STFC (PP/D000416/1).

\end{document}